\documentclass[aps,preprint]{revtex4}%
\usepackage{amsfonts}
\usepackage{amsmath}
\usepackage{amssymb}
\usepackage{graphicx}%
\providecommand{\U}[1]{\protect\rule{.1in}{.1in}}

\begin{document}
\title{ Magnetotransport of a periodically modulated graphene monolayer}
\preprint{ }
\author{R. Nasir}
\affiliation{Department of Physics,Quaid-i-Azam University, Islamabad 45320 Pakistan}
\author{K. Sabeeh$^{\ast}$}
\affiliation{Department of Physics,Quaid-i-Azam University, Islamabad 45320 Pakistan}
\author{M. Tahir}
\affiliation{Department of Physics, University of Sargodha, Sargodha $40100$, Pakistan.}
\keywords{one two three}
\pacs{PACS number}

\begin{abstract}
We have performed \ a detailed investigation of the electrical properties of a
graphene monolayer which is modulated by a weak one dimensional periodic
potential in the presence of a perpendicular magnetic field. The periodic
modulation broadens the Landau Levels into bands which oscillate with $B$. The
electronic conduction in this system can take place through either diffusive
scattering or collisional scattering off impurities. Both these contributions
to electronic transport are taken into account in this work. In addition to
the appearance of commensurability oscillations in both the collisional and
diffusive contributions, we find that Hall resistance also exhibits
commensurability oscillations. Furthermore, the period and amplitude of these
commensurability oscillations in the transport parameters and how they are
affected by temperature are also discussed in this work.

\end{abstract}
\volumeyear{year}
\volumenumber{number}
\issuenumber{number}
\eid{identifier}
\date[Date text]{date}
\received[Received text]{date}

\revised[Revised text]{date}

\accepted[Accepted text]{date}

\published[Published text]{date}

\startpage{1}
\endpage{2}
\maketitle

\section{INTRODUCTION}

Recent successful preparation of a single layer of graphene has generated a
lot of interest in this system as experimental and theoretical studies have
shown that the nature of quasiparticles in this two-dimensional system is very
different from those of conventional two-dimensional electron gas (2DEG)
systems realized in semiconductor heterostructures. Graphene has a honeycomb
lattice of carbon atoms. The quasiparticles in graphene have a band structure
in which electron and hole bands touch at two points in the Brillouin zone. At
these Dirac points the quasiparticles obey the massless Dirac equation. In
other words, they behave as massless, chiral Dirac Fermions leading to a
linear dispersion relation $E_{k}=\hbar v_{F}k$ (with the characteristic
velocity $v_{F}\simeq10^{6}m/s)$. This difference in the nature of the
quasiparticles in graphene from a conventional 2DEG has given rise to a host
of new and unusual phenomena such as the anomalous quantum Hall
effect\cite{1,2} with profound effects on transport in these systems. The
transport properties of graphene are currently being explored in the presence
of nonuniform potentials, such as in $p-n$ junctions\cite{3}, as well as in
periodic potentials. Effects of periodic potential on electron transport in 2D
electron systems has been the subject of continued interest, where electrical
modulation of the 2D system can be carried out by depositing an array of
parallel metallic strips on the surface or through two interfering laser
beams\cite{4}. More recently in graphene, electrostatic\cite{5} and
magnetic\cite{6} periodic potentials have been shown to modulate its
electronic structure in unique ways leading to fascinating physics and
possible applications. Periodic potentials are induced in graphene by
interaction with a substrate\cite{7} or controlled adatom deposition\cite{8}.
In addition, it was shown that periodic ripples in suspended graphene also
induces a periodic potential in a perpendicular electric field\cite{9}.
Epitaxial growth of graphene on top of a prepatterned substrate is also a
possible route to modulation of the potential seen by the electrons. In this
work, we complement these recent studies to discuss the effects of a weak
electric modulation on the electrical conductivity in a graphene monolayer
subjected to an external magnetic field perpendicular to the graphene plane.
Electric modulation introduces a new length scale, period of modulation, in
the system giving rise to interesting physical effects on the transport
response. Commensurabiliy (Weiss) oscillations, in addition to Shubnikov de
Hass (SdH) oscillations, are found to occur as a result of commensurability of
the electron cyclotron diameter at the Fermi energy and the period of the
electric modulation. In \cite{10} , on the same subject, diffusive
contribution to magnetoconductivity was considered whereas in the present work
we determine collisional and Hall contibutions as well. This makes this paper
a complete study of electric modulation induced effects on electrical
conductivities/resistivities in a graphene monolayer in the presence of a
magnetic field.

In the next section, we present the formulation of the problem and derive
expressions for electrical conductivities in a graphene monolayer. In section
III, results of numerical work are presented and discussed, followed by the
conclusions in section IV.

\section{FORMULATION}

We consider a graphene sheet in the $x-y$ plane. The magnectic field
$\mathbf{B}$ is applied along the $z-$ direction. The system is also subjected
to a 1D weak periodic modulation $U(x)$ in the $x-$ direction. The one
electron Hamiltonian reads%
\begin{equation}
H=v_{F}\mathbf{\sigma}.(p+e\mathbf{A})+U(x) \label{1}%
\end{equation}
where $p$ is the momentum operator, $\mathbf{\sigma}=\left\{  \sigma
_{x},\sigma_{y}\right\}  $ are Pauli matrices and $v_{F}(\sim10^{6}m/s)$
characterizes the electron velocity in graphene. In the absence of modulation,
i.e. for $U(x)=0$ and for the vector potential chosen in the Landau guage
$A=(0,Bx,0)$, the normalized eigenfunctions of Eq. (\ref{1}) are given by
$\frac{e^{ik_{y}y}}{\sqrt{2L_{y}l}}\binom{-i\phi_{n}(\frac{x+x_{o}}{l})}%
{\phi_{n-1}(\frac{x+x_{o}}{l})}$ where $\phi_{n}(x)$ and $\phi_{n-1}(x)$ are
the harmonic oscillator wavefunctions centred at $x_{o}=l^{2}k_{y}$. $n$ is
the Landau level index, $l=\sqrt{\frac{\hslash}{eB}}$ the magnetic length and
$L_{y}$ the length of 2D graphene system in the $y$ direction. The
corresponding eigenvalue is $E_{n}=\hslash\omega_{g}\sqrt{n}$ where
$\omega_{g}=v_{F}\sqrt{2eB/\hbar}=v_{F}\sqrt{2}/l$.

The modulation potential is approximated by the first Fourier component of the
periodic potential $U(x)=V_{o}\cos Kx$ where $K=2\pi/a$, $a$ is the period of
modulation and $V_{o}$ is the constant modulation amplitude. This potential
lifts the degeneracy of Landau Levels (LLs) and the energy becomes dependent
on the position$\ x_{o}$ of the guiding centre. Thus energy eigenvalues for
weak modulation ($V_{o}\ll E_{F}$), using first order perturbation theory, are%
\begin{equation}
E_{n,k_{y}}=E_{n}+V_{n,B}\cos Kx_{o} \label{2}%
\end{equation}
where $V_{n,B}=\frac{V_{o}}{2}e^{-u/2}\left[  L_{n}\left(  u\right)
+L_{n-1}\left(  u\right)  \right]  $ with $L_{n}\left(  u\right)  $ ,
$L_{n-1}\left(  u\right)  $ the Laguerre polynomials and $u=K^{2}l^{2}/2$. We
note that the electric modulation induced broadening of the energy spectrum is
nonuniform. The Landau bandwidth $\sim V_{n,B}$ oscillates as a function of
$n$ since $L_{n}\left(  u\right)  $ are oscillatory functions of index $n$.
$V_{n,B}$ at the Fermi energy can be approximated, using an asymptotic
expression for $n\gg1$ appropriate for low magnetic-field range relevant to
the present study, as%

\begin{equation}
V_{B}=V_{o}\sqrt{\frac{2}{\pi KR_{c}}}\cos(KR_{c}-\frac{\pi}{4}) \label{3}%
\end{equation}
where $R_{c}=k_{F}l^{2}$ is the classical cyclotron orbit, $k_{F}=\sqrt{2\pi
n_{e}}$ and $n_{e}$ is the electron number density. The above expression shows
that $V_{B}$ oscillates with $B,$ through $R_{c},$ and the width of Landau
bands $2\mid V_{B}\mid$ becomes maximum at%
\begin{equation}
\frac{2R_{c}}{a}=i+\frac{1}{4}\ \ \ (i=1,2,3,...) \label{4}%
\end{equation}
and vanishes at%
\begin{equation}
\frac{2R_{c}}{a}=i-\frac{1}{4}\ \ \ (i=1,2,3,...). \label{5}%
\end{equation}
which is termed the flat band condition. The oscillations of the Landau
bandwith is the origin of the commensurability (Weiss) oscillations and, at
the same time, are responsible for the modulation of the amplitude and the
phase of the Shubnikov-de Hass (SdH) oscillations.

To calculate the electrical conductivity in the presence of weak modulation we
use Kubo formula \cite{11}. The diffusive contribution to conductivity which
arises due to the scattering induced migration of the \ Larmor circle center
has already been determined for a graphene monolayer in \cite{10}. Our focus,
in this work, will be the calculation of the collisional contribution to the
conductivity and the the Hall conductivity.

\subsubsection{\textbf{COLLISIONAL\ CONDUCTIVITY:}}

To obtain collisional contribution to conductivity, we assume that electrons
are elastically scattered by randomly distributed charge impurities as it has
been shown that charged impurities play a key role in the transport properties
of graphene near the Dirac point\cite{17,18}. This type of scattering is
dominant at low temperature. The collisional conductivity when spin degeneracy
is considered is given by \cite{11}%
\begin{equation}
\sigma_{xx}^{\operatorname{col}}=\frac{\beta e^{2}}{\Omega}\underset{\xi
,\xi^{\prime}}{%
{\displaystyle\sum}
}f_{\xi}(1-f_{\xi^{\prime}})W_{\xi\xi^{\prime}}(\alpha_{x}^{\xi}-\alpha
_{x}^{\xi^{\prime}})^{2} \label{6}%
\end{equation}
where $f_{\xi}=[\exp(\frac{E_{\xi}-\mu}{k_{B}T}+1)]^{-1}$ is the Fermi Dirac
distribution function with $f_{\xi}=$ $f_{\xi^{\prime}}$ for elastic
scattering, $k_{B}$ is the Boltzmann constant and $\mu$ the chemical
potential. $W_{\xi\xi^{\prime}}$ is the transmission rate between the
one-electron states $\left\vert \xi\right\rangle $ and $\left\vert \xi
^{\prime}\right\rangle $, $\Omega$ the volume of the system, $e$ the electron
charge, $\tau(E)$ the relaxation time and $\alpha_{x}^{\xi}=\left\langle
\xi\right\vert r_{x}$ $\left\vert \xi\right\rangle $ the mean value of the $x$
component of the position operator when the electron is in state $\left\vert
\xi\right\rangle $.

Collisional conductivity arises as a result of migration of the cyclotron
orbit due to scattering by charge impurities. The scattering rate $W_{\xi
\xi^{\prime}}$ is given by%
\begin{equation}
W_{\xi\xi^{\prime}}=\underset{q}{%
{\displaystyle\sum}
}\left\vert U_{q}\right\vert ^{2}\left\vert \left\langle \xi\right\vert
e^{iq.(r-R)}\left\vert \xi^{\prime}\right\rangle \right\vert ^{2}\delta
(E_{\xi}-E_{\xi^{\prime}}). \label{7}%
\end{equation}
The Fourier transform of the screened impurity potential is $U_{q}=2\pi
e^{2}/\varepsilon\sqrt{q^{2}+k_{s}^{2}},$ where $r$ and $R$ are the position
of electron and of impurity respectively; $k_{s}$ is the screening wave
vector, $\varepsilon$ is the dielectric constant of the material. By
performing an average over random distribution of impurities, ($N_{I}\equiv$
impurity density), the contribution of the unperturbed part of the
wavefunction,$\left\vert \xi\right\rangle \equiv\left\vert n,k_{y}%
\right\rangle $, to the scattering rate is%
\begin{equation}
W_{\xi\xi^{%
\acute{}%
}}^{\left(  \circ\right)  }=\frac{2\pi N_{I}}{A_{\circ}\hslash}\underset
{q}{\sum}\left\vert U_{q}\right\vert ^{2}\left\vert \left\langle
n,k_{y}\right\vert e^{iq.(r-R)}\left\vert n^{\prime},k_{y}^{\prime
}\right\rangle \right\vert ^{2}\delta(E_{n,k_{y}}-E_{n^{\prime},k_{y}^{\prime
}}) \label{8}%
\end{equation}
with%
\begin{equation}
\left\vert \left\langle n,k_{y}\right\vert e^{iq.(r-R)}\left\vert n^{\prime
},k_{y}^{\prime}\right\rangle \right\vert ^{2}=\frac{1}{4}\left[
J_{n,n^{\prime}}(\gamma)+J_{n-1,n^{\prime}-1}(\gamma)\right]  \delta
_{k_{y}-k_{y}^{\prime},q_{y}} \label{9}%
\end{equation}
and%
\begin{equation}
\left\vert J_{n,n^{\prime}}(\gamma)\right\vert ^{2}=\frac{n!}{n^{\prime}%
!}e^{-\gamma}\gamma^{n-n^{\prime}}\left[  L_{n^{\prime}}^{n-1}\left(
\gamma\right)  \right]  ^{2};n^{\prime}\leq n. \label{10}%
\end{equation}
Here $A_{\circ}=L_{x}L_{y}$ is the area of the graphene monolayer and
$\gamma=l^{2}(q_{x}^{2}+q_{y}^{2})/2=\frac{q_{\perp}^{2}l^{2}}{2}$ with
$q_{\perp}^{2}=(q_{x}^{2}+q_{y}^{2})$. Inserting Eq. (\ref{8}) in Eq.
(\ref{6}) we obtain%
\begin{equation}
\sigma_{xx}^{\operatorname{col}}=\frac{e^{2}\beta l^{4}}{A_{\circ}}\frac{2\pi
N_{I}}{A_{\circ}\hslash}\underset{n,k_{y}}{\sum}\underset{n^{\prime},k_{y}%
}{\sum}\underset{q}{\sum}\left\vert U_{q}\right\vert ^{2}\frac{1}{4}\left[
J_{n,n^{\prime}}(\gamma)+J_{n-1,n^{\prime}-1}(\gamma)\right]  ^{2}%
q_{y}f_{n,k_{y}}(1-f_{n,k_{y}})\delta(E_{n,k_{y}}-E_{n^{\prime},k_{y}})
\label{11}%
\end{equation}
with $f_{n,k_{y}}\equiv f(E_{n,k_{y}})$, the Fermi Dirac distribution
function. Taking $\underset{q}{\sum}\rightarrow\frac{A_{\circ}}{4\pi^{2}l^{2}%
}\overset{2\pi}{\underset{0}{\int}}d\varphi\overset{\infty}{\underset{0}{\int
}}d\gamma$ and $q_{y}=q_{\perp}\sin\varphi$, $\left\vert U_{q}\right\vert
^{2}\sim\left\vert U_{\circ}\right\vert ^{2}$ in Eq. (\ref{11}),we obtain%
\begin{equation}
\sigma_{xx}^{\operatorname{col}}=\frac{e^{2}\beta N_{I}}{A_{\circ}\hslash
}\left\vert U_{\circ}\right\vert ^{2}\underset{n,n^{\prime},k_{y}}{\sum
}f_{n,k_{y}}(1-f_{n,k_{y}})\overset{\infty}{\underset{0}{\int}}\frac{1}%
{4}\gamma\left[  J_{n,n^{\prime}}(\gamma)+J_{n-1,n^{\prime}-1}(\gamma)\right]
^{2}d\gamma\delta(E_{n,k_{y}}-E_{n^{\prime},k_{y}}). \label{12}%
\end{equation}
Using the following integral identity \cite{11,16}:%
\begin{equation}
\overset{\infty}{\underset{0}{\int}}\gamma\left[  J_{n,n^{\prime}}%
(\gamma)\right]  ^{2}d\gamma=\overset{\infty}{\underset{0}{\int}}\gamma
e^{-\gamma}\left[  L_{n}(\gamma)\right]  ^{2}d\gamma=(2n+1) \label{13}%
\end{equation}
where for $n=n^{\prime}$, $\left[  J_{n,n^{\prime}}(\gamma)\right]
^{2}=e^{-\gamma}\left[  L_{n}(\gamma)\right]  ^{2}$ with the result%
\begin{equation}
\overset{\infty}{\underset{0}{\int}}\gamma\left[  J_{n-1.n^{\prime}-1}%
(\gamma)\right]  ^{2}d\gamma=\overset{\infty}{\underset{0}{\int}}\gamma
e^{-\gamma}\left[  L_{n-1}(\gamma)\right]  ^{2}d\gamma=(2n-1) \label{14}%
\end{equation}%
\begin{equation}
\overset{\infty}{\underset{0}{\int}}\gamma J_{n,n^{\prime}}(\gamma
)J_{n-1,n^{\prime}-1}(\gamma)d\gamma=\overset{\infty}{\underset{0}{\int}%
}\gamma e^{-\gamma}\left[  L_{n}(\gamma)\right]  \left[  L_{n-1}%
(\gamma)\right]  d\gamma=0. \label{15}%
\end{equation}
Finally, replacing the $\delta$ function by a Lorentzian of zero shift and
constant width $\Gamma$, $\underset{k_{y}}{\sum}\rightarrow\frac{L_{y}}{2\pi
}\overset{a/l^{2}}{\underset{0}{\int}dk_{y}}$, $A_{\circ}\rightarrow
L_{x}L_{y},$ and performing the sum on $n^{\prime}$, keeping only the dominant
term $n^{\prime}=n$ in Eq. (\ref{12}), we obtain the following result%
\begin{equation}
\sigma_{xx}^{\operatorname{col}}\approx\frac{e^{2}}{h}\frac{N_{I}U_{\circ}%
^{2}}{\pi a\Gamma}\underset{n=0}{\overset{\infty}{\sum}}n\overset{a/l^{2}%
}{\underset{0}{\int}}dk_{y}\beta f_{n,k_{y}}(1-f_{n,k_{y}}). \label{16}%
\end{equation}

\subsubsection{\textbf{DIFFUSIVE\ CONDUCTIVITY:}}

For completeness, we also present the result for diffusive conductivity which
was determined in\cite{10},%
\begin{equation}
\sigma_{yy}^{diff}=2\pi^{2}\frac{e^{2}}{h}\frac{V_{o}^{2}\tau\beta}{\hslash
}ue^{-u}\underset{n=0}{\overset{\infty}{\sum}}[\frac{-\partial f(E)}{\partial
E}]_{_{E=E_{n}}}[L_{n}(u)-L_{n-1}(u)]^{2} \label{17a}%
\end{equation}
where $\tau$ is the constant scattering time and $\frac{-\partial
f(E)}{\partial E}=\exp\beta(E-E_{F})/[\exp\beta(E-E_{F})+1]^{2}$. Now
$\sigma_{yy}=\sigma_{xx}^{\operatorname{col}}+\sigma_{yy}^{diff}$.

\subsubsection{\textbf{HALL CONDUCTIVITY:}}

The nondiagonal contribution to conductivity \cite{11} is given by%
\begin{equation}
\sigma_{yx}=\frac{2i\hslash e^{2}}{\Omega}\underset{\xi\neq\xi^{\prime}}{\sum
}f_{\xi}(1-f_{\xi^{\prime}})\left\langle \xi\right\vert v_{y}\left\vert
\xi^{\prime}\right\rangle \left\langle \xi^{\prime}\right\vert v_{x}\left\vert
\xi\right\rangle \frac{1-e^{\beta\left(  E_{\xi}-E_{\xi^{\prime}}\right)  }%
}{\left(  E_{\xi}-E_{\xi^{\prime}}\right)  ^{2}}. \label{17}%
\end{equation}
Since $f_{\xi}(1-f_{\xi^{\prime}})(1-e^{\beta\left(  E_{\xi}-E_{\xi^{\prime}%
}\right)  })=f_{\xi^{\prime}}(1-f_{\xi})$ and $\Omega\rightarrow A_{\circ
}\equiv L_{x}L_{y},$ we obtain%
\begin{equation}
\sigma_{yx}=\frac{2i\hslash e^{2}}{\Omega}\underset{\xi\neq\xi^{\prime}}{\sum
}f_{\xi^{\prime}}(1-f_{\xi})\frac{\left\langle \xi\right\vert v_{y}\left\vert
\xi^{\prime}\right\rangle \left\langle \xi^{\prime}\right\vert v_{x}\left\vert
\xi\right\rangle }{\left(  E_{\xi}-E_{\xi^{\prime}}\right)  ^{2}}. \label{18}%
\end{equation}
Since the $x$ and $y$ components \ of velocity operator are $v_{x}%
=\frac{\partial H_{\circ}}{\partial p_{x}}$ and $v_{y}=\frac{\partial
H_{\circ}}{\partial p_{y}}$ when $H_{\circ}=v_{F}\sigma.(p+eA)$. Therefore,
$v_{x}=v_{F}\sigma_{x}$ and $v_{y}=v_{F}\sigma_{y}$. Hence%
\begin{equation}
\left\langle \xi^{\prime}\right\vert v_{x}\left\vert \xi\right\rangle
=\left\langle n^{\prime},k_{y}\right\vert v_{x}\left\vert n,k_{y}\right\rangle
=-iv_{F} \label{19}%
\end{equation}
and%
\begin{equation}
\left\langle \xi\right\vert v_{y}\left\vert \xi^{\prime}\right\rangle
=\left\langle n,k_{y}\right\vert v_{y}\left\vert n^{\prime},k_{y}\right\rangle
=v_{F}. \label{20}%
\end{equation}
Substituting the values of the matrix elements of velocity in Eq. (\ref{18})
yields%
\begin{equation}
\sigma_{yx}=\frac{2\hslash e%
{{}^2}%
v_{F}%
{{}^2}%
}{L_{x}Ly}\underset{\xi\neq\xi^{\prime}}{\sum}\frac{f_{\xi^{\prime}}(1-f_{\xi
})}{\left(  E_{\xi}-E_{\xi^{\prime}}\right)  ^{2}}. \label{21}%
\end{equation}
Since $E_{\xi}\equiv E_{n,k_{y}}=E_{n}+V_{n,B}\cos Kx_{o}$ where
$E_{n}=\hslash\omega_{g}\sqrt{n}$ and $V_{n,B}=\frac{V_{o}}{2}e^{-u/2}\left[
L_{n}\left(  u\right)  +L_{n-1}\left(  u\right)  \right]  $ we obtain%
\begin{equation}
\left(  E_{\xi}-E_{\xi^{\prime}}\right)  ^{2}=\hslash^{2}\omega_{g}^{2}\left[
\sqrt{n+1}-\sqrt{n}+\lambda_{n}\cos Kx_{\circ}\right]  ^{2} \label{22}%
\end{equation}
where%
\begin{equation}
\lambda_{n}=\frac{V_{\circ}}{2\hslash\omega_{g}}e^{-u/2}\left(  L_{n+1}\left(
u\right)  -L_{n-1}\left(  u\right)  \right)  . \label{23}%
\end{equation}
Substituting Eq. (\ref{22}) in Eq. (\ref{21}) we obtain the Hall conductivity
in graphene as%
\begin{equation}
\sigma_{yx}=\frac{e^{2}}{h}\frac{l^{2}}{a}\underset{n=0}{\overset{\infty}%
{\sum}}\overset{a/l^{2}}{\underset{0}{\int}}dk_{y}\frac{f_{n,k_{y}%
}-f_{n+1,k_{y}}}{\left[  \sqrt{n+1}-\sqrt{n}+\lambda_{n}\cos Kx_{\circ
}\right]  ^{2}} \label{24}%
\end{equation}
Elements of the resistivity tensor $\rho_{\mu\nu}$($\mu$,$\nu$=$x$,$y$) can be
determined from those of the conductivity tensor $\sigma_{\mu\nu}$, obtained
above, using the expressions: $\rho_{xx}=$ $\sigma_{yy}$ $/S$, $\rho_{yy}=$
$\sigma_{xx}$ $/S$ and $\rho_{xy}=$ $-\sigma_{yx}$ $/S$ where $S=$
$\sigma_{xx}$ $\sigma_{yy}-$ $\sigma_{xy}$ $\sigma_{yx}$ with $S\approx$
$\sigma_{xy}^{2}=n_{e}^{2}e^{2}/B^{2}$.\bigskip

\section{RESULTS\ AND\ DISCUSSION}

The above expressions for the (collisional, diffusive and Hall)
conductivities, Eqs. (\ref{16}), (17) and (\ref{24}) are the principal results
of this work. The integrals appearing in these equations are evaluated
numerically and the results are presented in Figure (1a) at temperature $T=2$
$K$ for a graphene monolayer with electron density $n_{e}=3.0\times
10^{11}cm^{-2}$, electric modulation strength $V_{o}=0.5meV$ with period
$a=350nm$. In addition, the following parameters were employed \cite{18,19,20}%
: $\tau=4\times10^{-13}$ $s$, $\Gamma=0.4meV$, impurity density $N_{I}%
=2.5\times10^{11}cm^{-2}$ and $\varepsilon=3.9$ (using SiO$_{2}$ as the
substrate material). We observe that SdH oscillations are visible in
collisional conductivity $\sigma_{xx}$ whereas the Hall conductivity
$\sigma_{yx}$ decreases with increasing magnetic field, $B$. Furthermore,
Weiss oscillations superimposed on SdH oscillations are seen in $\sigma_{yy}.$
To highlight the effects of modulation, we also calculate the correction to
the conductivity (change in conductivity) as a result of modulation which is
expressed as $\Delta\sigma_{\mu\nu}=\sigma_{\mu\nu}(V_{o})-\sigma_{\mu\nu
}(V_{o}=0)$ and is shown in Figure (1b). Electric modulation acting on the
system results in a positive contribution to $\Delta\sigma_{yy}$ and a
negative contribution to $\Delta\sigma_{xx}\ $whereas $\Delta\sigma_{yx}$
oscillates around zero. We find that $\Delta\sigma_{yy}\gg\Delta\sigma_{xx}$,
which is a consequence of the fact that $\Delta\sigma_{xx}$ has only
collisional contribution, while $\Delta\sigma_{yy}$, in addition to the
collisional part, has contributions due to band conduction which are much
larger. It is also seen that the oscillations in $\Delta\sigma_{xx}$ and
$\Delta\sigma_{yy}$ are $180^{o}$ out of phase. To determine the effects of
temperature on magnetoconductivities, comparison of conductivities and
corrections to the conductivities at two different temperatures $T=2K$ (solid
curve) and $T=6K$ (broken curve) are presented in Figures (2) and (3)
respectively. $\Delta\sigma_{xx}$ shows strong temperature dependence which is
a clear signature that SdH oscillations are dominant here. Oscillations in
$\Delta\sigma_{yy}$ show comparitively weaker dependence on temperature as
Weiss oscillations, that are weakly dependent on temperature, play a more
significant role in $\sigma_{yy}.$ Furthermore, Weiss oscillations are also
seen in $\Delta\sigma_{yx}$ and they are weakly sensitive to temperature a low
magnetic fields (that is when $B<0.188T$). In graphene system, the value of
$B$ defining the boundary between SdH and Weiss oscillations is quite low (it
lies between $0.1$ and $0.15Tesla$). For smaller values of $B,$ the amplitude
of Weiss oscillations remain essentially the same at various temperatures.
When $B$ is large, SdH oscillations dominate and the amplitude of oscillations
gets reduced considerably at comparatively higher temperatures. However,
oscillatory phenomenon still persists.

It can be seen from Figure (1a), (2a) and (2b) that amplitude of SdH
oscillations remains large at those values of the magnetic field where the
flat band condition is satisfied i.e at
$B(Tesla)=0.6897,0.2956,0.1881,0.1379,0.1089...$ when $i=1,2,3,4...$in
Eq.(\ref{5}) while is supressed at the maximum bandwidth/broad band condition,
i.e at $B(Tesla)=0.4138,0.2299,0.1592,0.1217,0.0985,...$ for $i=1,2,3,4,...$
in Eq.(\ref{4}). Furthermore, zeros in $\Delta\sigma_{\mu\nu}$ appear in close
agreement with values predicted from the flat band condition. The amplitude of
$\Delta\sigma_{xx}$ and $\Delta\sigma_{yy}$ becomes maximum at the broad band
condition(as seen in Figure (3)), whereas the amplitude of $\Delta\sigma_{yx}$
crosses the zero level at the broad band conditon and than a phase change of
amplitude occurs.

Components of the resistivity tensor $\rho_{\mu\nu}$ have also been computed
and shown in Figure (4a) as a function of $B$ for $T=2K$ (solid curve) and
$6K$ (broken curve) respectively. The correction (change) in $\rho_{\mu\nu}$
due to the modulation is shown in Figure (4b). To verify our results, we
compare them in the absence of modulation with the unmodulated experimental
results presented in \cite{14}. In order to carry this out, we note that the
number density $n_{e\text{ }}$is related to the gate voltage ($V_{g}$) through
the relationship \cite{13}$\ n_{e}=\epsilon_{o}\epsilon V_{g}/te$, where
$\epsilon_{o}$ and $\epsilon$ are the permitivities for free space and the
dielectric constant of graphene, respectively. $e$ is the electron charge and
$t$ the thickness of the sample. It yields $V_{g}=4.8V$ for $n_{e}%
=3.0\times10^{11}cm^{-2}$. We find that the results for magnetoresistivities
obtained in this work are in good agreement with the values given in reference
\cite{14} for the unmodulated case at $V_{g}=4.8V$.

We observe in Figure (4), that the dominant effect of Weiss oscillations
appears in $\rho_{xx}$ as it is proportional to $\sigma_{yy}$ whereas the
amplitude of oscillations in $\rho_{yy}$ show a monotonic increase in ampitude
with magnetic field signifying dominance of SdH in $\rho_{yy}.$ In Figure (5),
we observe that the oscillations in $\triangle\rho_{xx}$ and $\triangle
\rho_{yy}$ are out of phase and the amplitude of the oscillation in
$\triangle\rho_{xx}$ is greater than the amplitude of oscillation in
$\triangle\rho_{yy}$. The out of phase character of the oscillations can be
understood by realizing that the conduction along the modulation direction,
which contributes to $\rho_{yy},$ occurs due to hopping between Landau states
and it is minimum when the density of states at the Fermi level is minimum.
Oscillations in $\rho_{xx}$ are much larger than those in $\rho_{yy}$ as a new
mechanism of conduction due to modulation contributes to $\rho_{xx}$. To
highlight temperature effects on the modulated system, we present in Figure
(6), corrections to magnetoresistivities at two different temperatures ($2K$,
solid curve and $6K$, broken curve). These results exhibit SdH oscillation
when $B$ becomes greater than $0.188T$ as seen in Figures (5) and (6). The
Weiss oscillations in $\triangle\rho_{xx}$ are in phase with those of
$\triangle\rho_{xy}$. From Figure (5c) one might infer that Hall resistivity
is not affected by modulation. This is not so, as even Hall resistivity
carries modulation effects and that is seen if we draw the slope of $\rho
_{xy}$ as a function of magnetic field (Figure (7)).

In order to quantatively analyze the results presented in the figures we
consider the density of states (DOS) of this system. At finite temperature,
the oscillatory part of resistivities ($\triangle\rho/\rho_{o}$) are
proportional to the oscillatory part of the density of states (DOS) at the
Fermi energy, $A(T/T_{c})\triangle D(E_{F})/D_{o}$ where $A(T/T_{c})=(\frac
{T}{T_{c}})/\sinh(\frac{T}{T_{c}}),$ $D_{o}$ is the DOS and $\rho_{o}$ is
resistivity in the absence of magnetic field, respectively\cite{15}. For not
too small magnetic fields ($B\gtrsim0.05T$ ), $\triangle\rho/\rho_{o}%
\simeq(\omega_{g}\tau)^{2}\triangle\sigma/\sigma_{o}$ to a good approximation,
where $\sigma_{o}=\frac{e^{2}v_{F}^{2}}{2}\tau D_{o}$ represents conductivity
at zero magnetic field and $\tau$ is the relaxation time. The analytic
expression for the density of states (DOS) of a graphene monolayer in the
presence of a magnetic field subjected to electric modulation has been derived
in the Appendix. The DOS at energy $E$ is given as%
\begin{align}
D(E,V_{B})  &  =D_{o}\left[  1+2\underset{k=1}{\overset{\infty}{\sum}}\frac
{1}{2\pi}\overset{2\pi}{\underset{0}{\int}}\cos[2\pi k(\varepsilon-v_{B}\cos
t)]dt\exp(-2\pi k\eta)\right] \label{25}\\
&  =D_{o}\left[  1+2\underset{k=1}{\overset{\infty}{\sum}}\cos(2\pi
k\varepsilon)J_{o}(2\pi kv_{B})\exp(-2\pi k\eta)\right] \nonumber
\end{align}
where $D_{o}=\frac{2E}{(\hbar\omega_{g})^{2}\pi l^{2}}=\frac{E}{\pi(\hbar
v_{F})^{2}}$, $\epsilon=(\frac{E}{\hbar\omega_{g}})^{2},\eta=\frac{\Gamma
E}{(\hbar\omega_{g})^{2}}$ and $v_{B}=\frac{2V_{B}E}{(\hbar\omega_{g})^{2}}.$
$J_{o}(x)$ is the Bessel function of order zero. Since $\exp(-2\pi k\eta)\ll1$
for weak magnetic fields, it is usually a good approximation to keep only the
$k=1$ term in the sum: $D(E)\simeq D_{o}+\Delta D_{1}(E)$ with%
\begin{equation}
\frac{\Delta D_{1}(E)}{D_{o}}=2\cos(2\pi\varepsilon)J_{o}(2\pi v_{B}%
)\exp(-2\pi\eta). \label{26}%
\end{equation}
To determine the effects of an external magnetic field on the
conductivities/resistivities of the system we consider Eq.(26). With a
decrease in $B$, $v_{B}$ oscillates periodically with respect to $1/B$ around
$v_{B}=0$, increasing its amplitude proportionaly to $1/\sqrt{B}$ [Eq.
(\ref{3})]. The function $J_{o}(2\pi v_{B})$ decreases from $1$ with an
increase of $\mid v_{B}\mid=0.3827\simeq3/8$ and than changes its sign.
Therefore the oscillations of $\Delta D_{1}(E)$ takes a minimum amplitude at
the maximum bandwith conditions while $\mid v_{B}\mid$ stays less than $3/8$;
it disappears when a maximum of $\mid v_{B}\mid$ touches at $\sim3/8$; it
reappears with an inverted sign for $\mid v_{B}\mid$ larger than $3/8$.
Therefore, if we assume that $\triangle\rho/[\rho_{o}A(T/T_{c})]$
$\propto\triangle D(E_{F})/D_{o}$ holds, we can find the position where
oscillations of $\triangle\rho/[\rho_{o}A(T/T_{c})]$ vanish . That occurs at
$\mid V_{B}\mid=0.19135(\hbar\omega_{g})^{2}/E_{F}$.

We can also find the period of oscillations in conductivities/resistivities
from Eq. (\ref{25}) as follows. We have $D(E,V_{B})\approx D_{o}\{1+2\cos
(2\pi\varepsilon)J_{o}(2\pi v_{B})\exp(-2\pi k\eta)\}\approx D_{o}%
\{1+2\cos(2\pi\varepsilon)(1-\pi^{2}v_{B}^{2})\exp(-2\pi k\eta)\}$. Since
$v_{B}^{2}\propto\cos^{2}(KR_{c}-\frac{\pi}{4})$. The period of oscillation
can be estimated by equating the increment of the cosine argument with $\pi$,%
\begin{equation}
K\Delta(R_{c})=\pi, \label{27a}%
\end{equation}
which leads to%
\begin{equation}
\Delta\left(  \frac{1}{B}\right)  =\left(  \frac{e}{2\sqrt{2\pi}\hbar}\right)
\frac{a}{\sqrt{n_{e}}}. \label{27}%
\end{equation}
In our work ($\ n_{e}=3.0\times10^{11}cm^{-2}$ and $a=350nm$), therefore the
period of oscillations comes out to be 1.933$T$ $^{-1}$ which is in good
agreement with the results shown in the figures.

Damping of these oscillations with temperature can also be discussed. In Ref.
\cite{10}, the temperature scale for damping of Weiss oscillations is given by
$K_{B}T_{c}^{Weiss}=b\hbar v_{F}/4\pi^{2}a$ where $b=(a/l)^{2}$ and
$v_{F}=\frac{\omega_{g}l}{\sqrt{2}},$ whence the result%
\begin{equation}
K_{B}T_{c}^{Weiss}=\frac{a\hbar v_{F}}{4\pi^{2}l^{2}}=\frac{\hbar\omega_{g}%
}{2\pi^{2}}\left(  \frac{a}{2\sqrt{2}l}\right)  . \label{28a}%
\end{equation}
To determine the damping temperature for SdH oscillations we, following
Ref.\cite{10} and \cite{11}, use asymptotic expression for
magnetoconductivity. For this, we use DOS (Eq. \ref{25}),%
\begin{equation}
D(E)=\frac{2E}{(\hbar\omega_{g})^{2}\pi l^{2}}\left[  1+2\exp(-2\pi\eta
)\cos(2\pi\frac{E^{2}}{(\hbar\omega_{g})^{2}})+...\right]  . \label{28b1}%
\end{equation}
In the asymptotic limit of weak magnetic fields when many filled Landau levels
occur, we take $L_{n}\approx$ $L_{n-1}$ and replace $e^{-u/2}L_{n}$ by
$1/\sqrt{\pi\sqrt{nu}}\cos(2\sqrt{nu}-\pi/4)$ and inserting the continuum
approximation $\underset{n=0}{\overset{\infty}{\sum}}\rightarrow
\overset{\infty}{\underset{0}{\int}}dED(E)$ in Eq. \ref{17a}, we obtain the
following result%
\begin{equation}
\frac{\sigma_{yy}^{diff}}{\sigma_{o}}=\frac{4\sqrt{2}\pi^{2}l}{a}\frac
{V_{o}^{2}}{E_{F}(\hbar\omega_{g})}\left[  F+2\exp(-2\pi\eta)A(T/T_{c}%
^{SdH})\cos\left(  2\pi\frac{E_{F}^{2}}{(\hbar\omega_{g})^{2}}\right)
\cos^{2}\left(  \sqrt{2}Kl\frac{E_{F}}{\hbar\omega_{g}}-\frac{\pi}{4}\right)
\right]  \label{28b2}%
\end{equation}
where $F=\frac{1}{2}\left[  1-A(T/T_{c}^{Weiss})+2A(T/T_{c}^{Weiss})\cos
^{2}\left(  \sqrt{2}Kl\frac{E_{F}}{\hbar\omega_{g}}-\frac{\pi}{4}\right)
\right]  $ is the contribution of Weiss oscillations and $A(T/T_{c}%
^{SdH})=[4\pi^{2}E_{F}K_{B}T/(\hbar\omega_{g})^{2}]/\sinh[4\pi^{2}E_{F}%
K_{B}T/(\hbar\omega_{g})^{2}]$ is the amplitude of the SdH oscillations.
Therefore, the characteristic temperature of SdH oscillations is given by%
\begin{equation}
K_{B}T_{c}^{SdH}=\frac{(\hbar\omega_{g})^{2}}{4\pi^{2}E_{F}}=\frac{\hbar
\omega_{g}}{2\pi^{2}}\left(  \frac{1}{\sqrt{2}k_{F}l}\right)  . \label{28b}%
\end{equation}
The amplitude of oscillations is given by $A=\frac{x}{\sinh(x)}$, where
$x=\frac{T}{T_{c}}$. The amplitude of Weiss oscillations at $B=0.3T$ are
$0.9993$ and $0.9938$ at $T=2K$ and $T=6K$, respectively The corresponding
amplitudes for SdH oscillations are $0.6878$ and $0.0882$. The SdH amplitude
decreases by $\sim87$ percent whereas the amplitude of Weiss ocillations
decreases by $\sim0.55$ percent for $4K$ change in temperature. In Figures
(2), (3), (5) and (6); the SdH amplitude decreases by $\sim77$ percent when
temperature is changed from $T=2K$ to $T=6K$, and it is in good agreement with
the results obtained from Eqs. (\ref{28a}) and (\ref{28b}). It is due to the
fact $K_{B}T_{c}^{Weiss}\gg K_{B}T_{c}^{SdH}$ that the Weiss oscillations are
more robust against temperatue changes.

Finally, we compare the results obtained for the conductivity/ resistivity of
graphene with those of a 2DEG given in \cite{11}. The characteristic damping
temperatures for Weiss and SdH oscillations in 2DEG are $K_{B}T_{2DEG}%
^{Weiss}=\frac{\hbar\omega_{c}}{2\pi^{2}}(\frac{ak_{F}}{2})$ and
$K_{B}T_{2DEG}^{SdH}=\frac{\hbar\omega_{c}}{2\pi^{2}},$ respectively. In
contrast, the corresponding damping temperatures in graphene are given by Eqs
(30) and (33). On comparing the two temperature scales, we find that the
damping temperatures of both oscillations in graphene are higher than that of
a 2DEG. The ratio is found to be $\frac{T_{c}}{T_{c,e}}=\frac{m^{\ast}v_{F}%
}{\hbar k_{F}}\approx4.2$; where $m^{\ast}$ is the electron mass in a 2DEG and
$T_{c,e}$ is the critical temperature of a 2DEG; which implies that a
comparatively higher temperature is required for damping of oscillations in
graphene. This is due to the higher Fermi velocity of Dirac electrons in
graphene compared to standard electrons in a 2DEG systems. It is evident from
the numerical results that both, Sdh and Weiss-type oscillations, are more
enhanced and more robust against temperature in graphene.

To conclude, we have investigated the effects of a weak periodic electric
modulation on the conductivity of a graphene monolayer subjected to a
perpendicular magnetic field. As a result of modulation a new length scale,
period of modulation, enters the system leading to commensurability
oscillations in the diffusive, collisional and Hall contributions to
conductivities/resistivities. These modulation induced effects on graphene
magnetotransport are discussed in detail in this work.

\section{Appendix}

Here we derive the expression for the density of states, Eq. (\ref{25}) in the
text. We consider monolayer graphene subjected to a uniform quantizing
magnetic field $\mathbf{B}=B\hat{z}$ in the presence of an additional weak
periodic modulation potential. The energy spectrum in the quasi classical
approximation, i.e. when many Landau bands are filled may be written as%
\begin{equation}
E_{n,x_{o}}=\sqrt{n}\hbar\omega_{g}+V_{n,B}\cos Kx_{o} \label{29}%
\end{equation}
where $V_{n,B}=\frac{V_{o}}{2}e^{-u/2}\left[  L_{n}\left(  u\right)
+L_{n-1}\left(  u\right)  \right]  $ with $L_{n}\left(  u\right)  $ ,
$L_{n-1}\left(  u\right)  $ the Laguerre polynomials and $u=K^{2}l^{2}/2$. For
large $n;$ $L_{n}\left(  u\right)  \approx L_{n-1}\left(  u\right)  $ and
$V_{n,B}=V_{o}e^{-u/2}L_{n}\left(  u\right)  $. Using the asymptotic
expression for the Laguerre polynomials\cite{16}; $e^{-u/2}L_{n}\left(
u\right)  \rightarrow\frac{1}{\sqrt{\pi\sqrt{nu}}}\cos(2\sqrt{nu}-\frac{\pi
}{4})$ and taking the continuum limit $n\rightarrow\frac{1}{2}(\frac{lE}%
{v_{F}\hbar})^{2}$, where $v_{F}=\omega_{g}l/\sqrt{2}$ we get
\begin{equation}
V_{n,B}=V_{0}\pi^{-1/2}\left(  \frac{1}{2}K^{2}l^{2}\frac{E}{\hbar\omega_{g}%
}\right)  ^{-1/4}\cos\left(  \sqrt{2}Kl\frac{E}{\hbar\omega_{g}}-\frac{\pi}%
{4}\right)
\end{equation}%
\begin{equation}
V_{n,B}=V_{0}\pi^{-1/2}\left(  \frac{1}{2}K^{2}l^{2}\frac{E}{\hbar\omega_{g}%
}\right)  ^{-1/4}\cos\left(  \sqrt{2}Kl\frac{E}{\hbar\omega_{g}}-\frac{\pi}%
{4}\right)  . \label{30}%
\end{equation}
To obtain a more general result which will lead to the result that we require
as a limiting case we consider impurity broadened Landau levels. The self
energy may be expressed as%
\begin{equation}
\Sigma^{-}(E)=\Gamma_{o}^{2}%
{\displaystyle\sum\limits_{n}}
{\displaystyle\int\limits_{0}^{a}}
\frac{dx_{o}}{a}\frac{1}{E-E_{n,x_{0}}-\Sigma^{-}(E)} \label{31}%
\end{equation}
which yields%
\begin{equation}
\Sigma^{-}(E)=%
{\displaystyle\int\limits_{0}^{a}}
\frac{dx_{o}}{a}\underset{-\infty}{\overset{\infty}{\sum}}\frac{\Gamma_{o}%
^{2}}{E-\Sigma^{-}(E)-V_{n,B}\cos Kx_{o}-\sqrt{n}\hbar\omega_{g}}. \label{32}%
\end{equation}
$\Gamma_{o}$ is the broadening of the levels due to the presence of
impurities. The density of states is related to the self energy through
\begin{equation}
D(E)=\operatorname{Im}\left[  \frac{\Sigma^{-}(E)}{\pi^{2}l^{2}\Gamma_{o}^{2}%
}\right]  . \label{33}%
\end{equation}
The residue theorem has been used to sum the series$\underset{-\infty
}{\overset{\infty}{\sum}}f(n)=-\{$Sumof residues of $\pi(\cot\pi n)f(n)$at all
poles of $f(n)\}$\cite{12}. Here $f(n)=\underset{-\infty}{\overset{\infty
}{\sum}}\frac{b}{c-d\sqrt{n}}$ with $b=\Gamma_{o}^{2}$, $c=E-\Sigma
^{-}(E)-V_{n,B}\cos Kx_{o}$ and $d=\hbar\omega_{g}$. The function $f(n)$ has a
pole at $c^{2}/d^{2}$ and the residue of ($\pi(\cot\pi n)f(n)$) at the pole is
$\frac{-2bc}{d^{2}}\pi\cot(\frac{\pi c^{2}}{d^{2}})$. Hence $\underset
{-\infty}{\overset{\infty}{\sum}}f(n)=\frac{2bc}{d^{2}}\pi\cot(\frac{\pi
c^{2}}{d^{2}})$ and we obtain%
\begin{equation}
\Sigma^{-}(E)=%
{\displaystyle\int\limits_{0}^{a}}
\frac{dx_{o}}{a}\frac{2\pi\Gamma_{o}^{2}(E-\Sigma^{-}(E)-V_{n,B}\cos Kx_{o}%
)}{(\hbar\omega_{g})^{2}}\cot\left(  \frac{\pi(E-\Sigma^{-}(E)-V_{n,B}\cos
Kx_{o})^{2}}{(\hbar\omega_{g})^{2}}\right)  \label{34}%
\end{equation}%
\[
\approx\frac{2\pi\Gamma_{o}^{2}E}{(\hbar\omega_{g})^{2}}%
{\displaystyle\int\limits_{0}^{a}}
\frac{dx_{o}}{a}\cot\left(  \frac{\pi E}{(\hbar\omega_{g})^{2}}[E-2\{\Sigma
^{-}(E)+V_{n,B}\cos(Kx_{o})\}]\right)  .
\]
Separating $\Sigma^{-}(\varepsilon)$ into real and imaginary parts
\begin{equation}
\Sigma^{-}(E)=\Delta(E)+i\frac{\Gamma(E)}{2}, \label{35}%
\end{equation}
Eq. (\ref{34}) takes the form%
\begin{equation}
\Delta(E)+i\frac{\Gamma(E)}{2}=\frac{2\pi\Gamma_{o}^{2}E}{(\hbar\omega
_{g})^{2}}%
{\displaystyle\int\limits_{0}^{a}}
\frac{dx_{o}}{a}\frac{\sin2u+i\sinh2v}{\cosh2v-\cos2u} \label{36}%
\end{equation}
where%
\begin{equation}
u=\frac{\pi E}{(\hbar\omega_{g})^{2}}[\varepsilon-2\{\Delta(E)+V_{n,B}%
\cos(Kx_{o})\}] \label{37}%
\end{equation}%
\begin{equation}
v=\frac{\pi\Gamma(E)E}{(\hbar\omega_{g})^{2}} \label{38}%
\end{equation}%
\begin{equation}
\operatorname{Im}\left[  \Sigma^{-}(E)\right]  =\frac{2\pi\Gamma_{o}^{2}%
E}{(\hbar\omega_{g})^{2}}%
{\displaystyle\int\limits_{0}^{a}}
\frac{dx_{o}}{a}\frac{\sinh2v}{\cosh2v-\cos2u}=\frac{2\pi\Gamma_{o}^{2}%
E}{(\hbar\omega_{g})^{2}}%
{\displaystyle\int\limits_{0}^{a}}
\frac{dx_{o}}{a}(1+2\underset{k=1}{\overset{\infty}{\sum}}\cos(2ku)\exp(-2kv).
\label{39}%
\end{equation}
If we define dimensionless variables $\varepsilon=(\frac{E}{\hbar\omega_{g}%
})^{2}$, $\eta=\frac{\Gamma E}{(\hbar\omega_{g})^{2}}$ and $v_{B}=\frac
{2V_{B}E}{(\hbar\omega_{g})^{2}}$ the density of states is obtained as%
\begin{equation}
D(E,V_{B})=D_{o}(E)\{1+2\underset{k=1}{\overset{\infty}{\sum}}%
{\displaystyle\int\limits_{0}^{a}}
\frac{dx_{o}}{a}\cos[2\pi k(\varepsilon-v_{B}\cos K_{o}x)]\exp(-2\pi k\eta)\}
\label{40}%
\end{equation}
where $D_{o}(E)=\frac{2E}{(\hbar\omega_{g})^{2}\pi l^{2}}$. Let $Kx_{o}=t$ in
the above expression results in
\begin{equation}
D(E,V_{B})=D_{o}(E)\{1+2\underset{k=1}{\overset{\infty}{\sum}}\frac{1}{2\pi
}\overset{2\pi}{\underset{0}{\int}}\cos[2\pi k(\varepsilon-v_{B}\cos
t)]dt\exp(-2\pi k\eta)\}. \label{41}%
\end{equation}
Solving the integeral yields%
\begin{equation}
D(E,V_{B})=D_{o}(E)\{1+2\underset{k=1}{\overset{\infty}{\sum}}\cos(2\pi
k\varepsilon)J_{o}(2\pi kv_{B})\exp(-2\pi k\eta)\}. \label{42}%
\end{equation}

$^{\ast}$Corresponding author: ksabeeh@qau.edu.pk

\end{document}